# Reliable Expanded Beam Connectors for Single Mode Optical Fiber Sensor Applications in Harsh Environment

Xavier Insou [a][c], Lionel Quétel [b], Sébastien Claudot [c], Monique Thual[a]

[a]Institut Foton, CNRS UMR6082, Université de Rennes 1, 6 rue de Kérampont, F-22305 Lannion, France

[b]IDIL Fibres Optiques, 21 rue Louis De Broglie 22300 Lannion, France

[c]Souriau-Sunbank Connection Technologies, RD323, 72470 Champagné, France

***Abstract***   We propose an expanded beam micro lens in standard single mode connectors, compliant with harsh environment connections. Low Insertion loss, high return loss and relaxed alignment tolerances compared with Single Mode Fiber (SMF) are demonstrated in C and O band. We give one example of harsh environment optical fiber sensor applications.

## I. INTRODUCTION

Connections in harsh environment are key elements in the composition of the measurement line in single mode optical fiber sensors. For example, in the ASHLEY European project (Avionics Systems Hosted on a distributed modular electronics Large scale dEmonstrator for multiple tYpes of aircraft), different optical sensors will be implanted in multiple types of aircrafts. One of the biggest challenges is often to couple light through the sensor device [1].

A solution to realize harsh single mode connectors is to increase the mode field diameter at the fiber output. This extension is generally obtained by using ball or hemispherical lenses at the end of the fibers [2]. This technique is interesting but not optimized in terms of insertion loss and return loss, it also needs a specific connector housing. In this article, we propose an innovative solution to increase the beam diameter with low insertion loss, low reflectance and good repeatability by using standard connectors. First, we give the concepts and principle of this solution, then we integrate it in standard connectors and measure the sensitivity to different contaminations. Finally, examples of optical fiber sensor demonstrators pointing out the interest of these expanded beam connectors are given.

## II. THEORETICAL CONCEPTS AND PRINCIPLE OF THE EXPANDED BEAM CONNECTORS

Due to their very small Mode Field Diameters (MFD), SMF are very sensitive to contaminants, laser power and alignment defaults either lateral $\Delta x$, axial $\Delta z$ or angular $\theta$ when injecting signal from one fiber to one another whatever the application as illustrated in Fig.1. Expanded beam micro-lenses are very useful to relax lateral and axial positioning tolerances as can be seen in Table 1, where defaults that lead to 1 dB excess loss are calculated with equations extracted from ref [3] for different kinds of expanded Gaussian beams whose MFD are 55, 113 and 340 µm, compared with SMF of 10.5 µm MFD at a wavelength of 1550 nm.

The lateral positioning tolerances are relaxed from 2.5 to 82 µm, the axial one from 57 to 59500 µm with a 340 µm MFD expanded beam compared with SMF. But the angular alignment decreases with expanded beam from 2.6 to 0.08°. Then a MFD around 55 to 110 µm is a good tradeoff to relax lateral and axial positioning tolerances without decreasing the angular ones compared with SMF.

*Corresponding author. e-mail : sclaudot@souriau.com

TABLE 1: TOLERANCES FOR 1 dB EXCESS LOSS @ λ=1550NM

| MFD (µm) | $\theta$ (°) | $\Delta x$ (µm) | $\Delta z$ (µm) |
|---|---|---|---|
| **SMF 10.5** | 2.6 | 2.5 | 57 |
| **55** | 0.5 | 13 | 1560 |
| **113** | 0.25 | 26 | 6250 |
| **340** | 0.08 | 82 | 59500 |

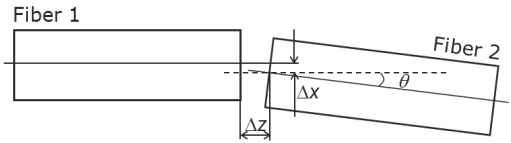

Fig. 1: Positioning defaults between two fibers

Moreover, expanded beams are less sensitive to high power and contaminants such as dust, oil and water for harsh environment applications.

A solution to achieve such an expanded beam at a SMF fiber output consists in splicing a parabolic lateral Graded Index Fiber (GIF) to a SMF and to cut the GIF, at a length $L_{GIopt}$ from the splicing, where the Gaussian beam size $2\omega$ is expanded at its maximum $2\omega_{max}$ as can be seen in Fig. 2. In this case the wave front is plane at the fiber output and the working distance $z_w$ is null. This configuration is appropriate for expanded beam connectors compliant with harsh conditions single mode fiber transmissions of at least 10 Gbit.s$^{-1}$ as already demonstrated in 55 µm MFD LC connectors as can be seen in Fig. 3. Coupling loss as low as 0.5 dB from O to C band have been achieved for Return Loss as high as 65 dB [4].

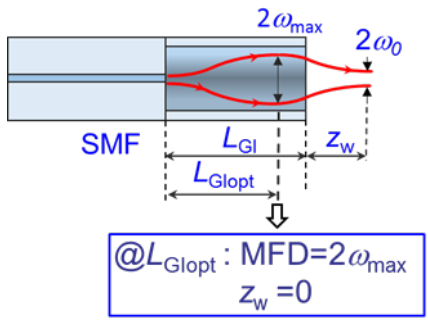

Fig. 2: Principle of the SMF GIF expanded beam micro lens

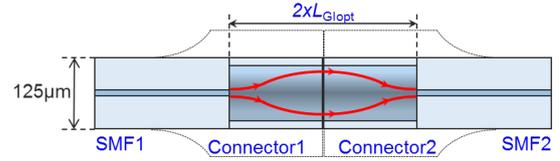

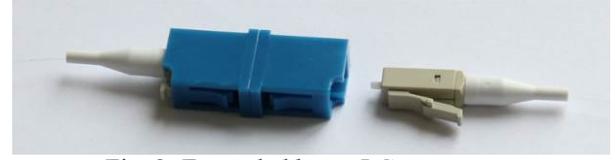

Fig. 3: Expanded beam LC connectors

Fig. 4 shows a photo of the 55 µm expanded beam at a SMF-GIF output compared with a SMF pointing out the expanded beam and the 125 µm outer diameter that remains constant, allowing the use of standard connectors (see Fig. 3).

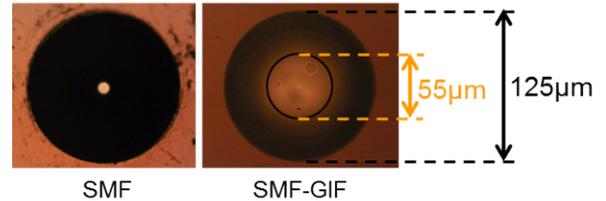

Fig. 4: Optical microscope view of a SMF-GIF output compared with a SMF when illuminated by visible light

We have also pointed out the best conditions to relax lateral and axial positioning tolerances without going down to critical angle tolerances and to achieve both high coupling efficiency and low reflectance with very tolerant fabrication process of this kind of expanded beam micro lenses [5].

Those expanded beam micro lenses can also be used in optical fiber sensor applications since functions can be added in between the fibers thanks to the relaxed axial and lateral positioning tolerances. For instance, 55 MFD expanded beam fibers can be separated from more than 1.5 mm with only 1 dB excess loss as can be seen in Table 1. Moreover, the micro lens can also be designed to achieve a wide range of MFD $2\omega_0$ and working distances $z_w$ depending on the $L_{GI}$ length and opto-geometric parameters of the GIF as can be seen in Fig. 5 where the MFD and working distance are plotted as a function of relative $L_{GI}$ length to $L_{GIopt}$ as developed in [4,5]. The maximum MFD is 55 µm for a null working distance but it can achieve a wide range of MFD comprised between 10.5 to 55 µm with working distances in a range of 0 up to 500 µm allowing a wide range of applications comprising optical fiber sensors.

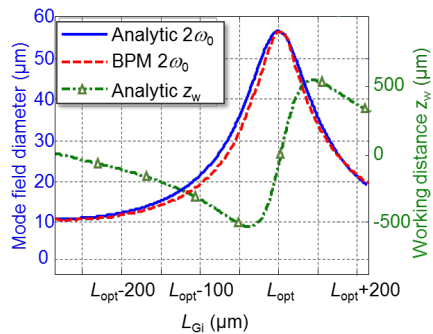

Fig. 5: Analytical and BPM calculation of MFD and working distance for a given GIF profile

## III. INTEGRATION OF EXPANDED BEAM FIBERS IN STANDARD CONNECTORS

These expanded beam fibers can easily be integrated in standard connectors such as FC, ST or even a multipoint one, as can be seen in Fig. 6 showing a picture of a connector with two expanded beam fibers. The fibers are glued in connector ferrules with a standard procedure. The ferrules are polished using conventional optic fiber polishing machines compatible with mass production. This operating mode authorizes repeatability in terms of Insertion Loss and Return Loss.

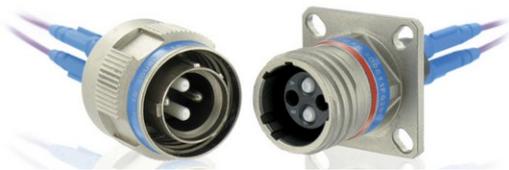

Fig. 6: Example of multipoint connectors with expanded beam fiber

## IV. CONTAMINATION EFFECT (OIL, WATER, DUST)

The robustness of the SMF GIF interface has been evaluated over three contaminants: oil (0.03ml droplet), water (0.03ml droplet) and dust particles in comparison with SMF and Multimode Fiber (MMF) (type OM3) as shown in Fig. 7. The insertion loss deviation compared with contaminant free connections in LC connectors is plotted for 10 mating cycles each time in Fig. 7. The physical contact technology is able to exclude the liquid from the interface due to the high pressure of the spring loading which leads to good optical results in water contamination. Sometimes some liquid residue is left, leading to higher losses for a single mode interface with an oil droplet for instance. With solid particle contamination, it is a matter of statistical particle distribution over the core area; when the remaining area for passing light becomes smaller, the insertion losses increase accordingly.

The results in Fig. 7 show that the SMF GIF is an interface with similar robustness compared to a MMF in the presence of contamination and better than SMF.

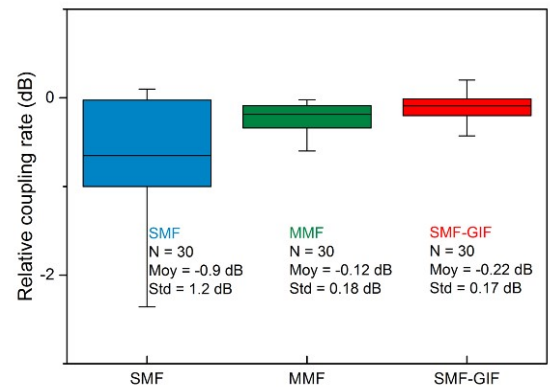

Fig. 7: Insertion loss deviation ($\Delta IL$) recorded over oil contamination (MIL-PRF-7808), water contamination and dust particle contamination (talc powder 10 microns) compared with contaminant free connections, for a SMF GIF LC connection, a SMF LC connection and a MMF LC connection, each time with 10 mating cycles

Such a behavior makes utilization of such connectors in harsh environments very useful especially in sensor applications. Moreover, thanks to the high misalignment tolerances as demonstrated in paragraph 2, these expanded beam connectors are less sensitive to thermal induced mechanical expansion.

## V. COMPATIBILITY OF EXPANDED BEAM CONNECTORS WITH OPTICAL SENSING APPLICATION

The testability of expanded beam connectors is essential to validate the global optical budget of a system. Fig. 8 represents the reflectometry testing done in Souriau labs on a link including three expanded beam connections. The peak of each connection is well defined and the IL/RL values fit with individual measurement.

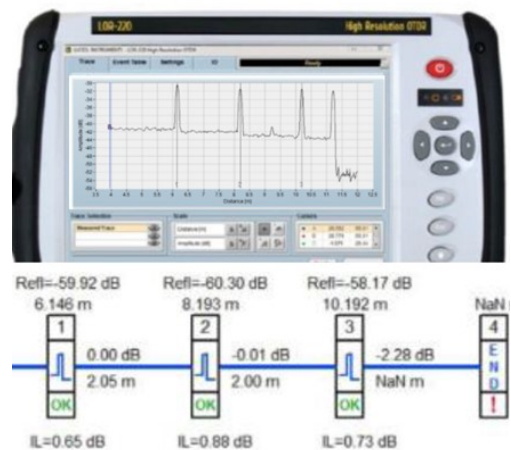

Fig. 8: Reflectometry testing of chained expanded beam connectors with a field production tool (Luciol LOR-220 OTDR)

The study of the dynamic behavior of a naval ship deck by using Fiber Bragg Gratings (FBGs) has been performed by IDIL Fibres Optiques. Arrays of FBGs sensors were used to measure structure elongation. The FBGs were connected to the interrogator with optical connectors. Fig. 9 represents the test setup and the spectrum of the 3 FBGs. No spectral deviation of the sensing signal was observed with expanded beam connection.

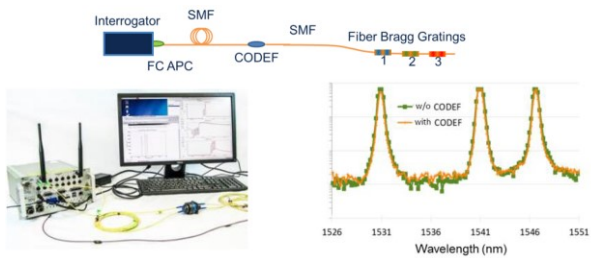

Fig. 9: Monitoring of a FBGs chain with and without expanded beam connection between FBGs and interrogator (Safibra FBGuard 1550)

Humidity, salt mist, oil and vibrations make the use of the proposed expanded beam connectors unavoidable for long term reliability in one such marine environment.

## VI. CONCLUSION

We have theoretically analyzed and experimentally demonstrated the use of innovative expanded beam fibers integrated in standard connectors for SM optical sensor applications with low insertion loss and low reflectance. The connectors authorize transmissions of at least 10 Gbits per second, can be used in a broad range of wavelengths (>200 nm) and should also support higher power than SMF (up to 100 W expected). Due to its specifications, such connectors can be used in a large range of applications including SHM (Structure Health Monitoring).

## ACKNOWLEDGMENT

This work has been done in the frame of the CODEF project ("Connectique Optique Durcie par Expansion de Faisceau") supported by a French government funding.